# A MATHEMATICAL DESCRIPTION OF THE DYNAMICS OF CORONAVIRUS DISEASE 2019 (COVID-19): A CASE STUDY OF BRAZIL

**Pedro V. Savi**
**Marcelo A. Savi**
**Beatriz Borges**

Center for Nonlinear Mechanics
COPPE – Department of Mechanical Engineering
Universidade Federal do Rio de Janeiro
21.941.972 – Rio de Janeiro – RJ, Brazil, P.O. Box 68.503
Email: pedrov.savi@gmail.com, savi@mecanica.coppe.ufrj.br, beatrizrbborges@gmail.com

**ABSTRACT**

This paper deals with the mathematical modeling and numerical simulations related to the coronavirus dynamics. A description is developed based on the framework of susceptible-exposed-infectious-removed model. Initially, a model verification is carried out calibrating system parameters with data from China, Italy, Iran and Brazil. Afterward, numerical simulations are performed to analyze different scenarios of COVID-19 in Brazil. Results show the importance of governmental and individual actions to control the number and the period of the critical situations related to the pandemic.

## 1. INTRODUCTION

Coronaviruses is related to illness that vary from a common cold to more severe diseases related to respiratory syndromes. Coronavirus disease 2019 (COVID-19) was discovered in 2019, the first time identified in humans. It is zoonotic, which means that it is transmitted among animals and humans. In January 21, 2019, World Health Organization (WHO) published the first Situation Report about the novel coronavirus (2019-nCoV). It announces to the world the origin of the COVID-19, reporting cases of pneumonia of unknown etiology detected in Wuhan City, Hubei Province of China. Afterward, the situation evolves to a huge global crisis with severe effects in Italy, Iran, Spain, South Korea and all over the world. In 11 March 2020, WHO declared that COVID-19 can be characterized as pandemic.

This dramatic situation points that all tools can be useful to define the best strategies for the public health system. In this regard, mathematical modeling is an interesting approach that can allow the evaluation of different scenarios, furnishing information for a proper support for health system decisions. In general, nonlinear dynamics of biological and biomedical systems is the objective of several researches that can be based on mathematical modeling or time series analysis (Savi, 2005). In particular, coronavirus propagation can be described by a mathematical model that allows the nonlinear dynamics analysis, representing different populations related to the phenomenon.

Literature presents some examples related to the dynamics of infectious diseases. Different kinds of models can be employed, essentially considering nonlinear governing equations. Rihan *et al.* (2018) described the dynamics of coronavirus infection in human, establishing interaction among human cells and the virus.

Chen *et al.* (2020) developed a mathematical model for calculating the transmissibility of the virus considering a simplified version of the bats-hosts-reservoir-people transmission model, defined as a reservoir-people model. Results follow the general trend of the initial propagation. Li *et al.* (2020) estimated characteristics of the epidemiologic time distribution, exploiting some pattern trends of transmission propagation. Riou & Althaus (2020) exploited the pattern of human-to-human transmission of novel coronavirus in Wuhan, China. Two key parameters are considered: basic reproduction number that defines the infectious propagation; and the individual variation in the number of secondary cases. Uncertainty quantification tools were employed to define the transmission patterns.

Susceptible-exposed-infectious-recovered (SEIR) models are an interesting approach to deal with the mathematical modeling of coronavirus transmission. Wu *et al.* (2020) investigated Wuhan – China case, evaluating nowcasting and forecasting domestic and international spread outbreak. Lin *et al.* (2020) proposed a model considering individual reaction, governmental action and emigration. The model is based on the original work of He *et al.* (2013) that proposed a model to describe the 1918 influenza.

The mathematical model proposed by Lin *et al.* (2020) seems to be capable to capture the general propagation of the novel coronavirus, being employed in this work to evaluate different scenarios of the propagation of coronavirus in Brazil. Initially, a model verification is carried out considering infected population evolution of China, Italy, Iran and Brazil. Afterward, Brazilian COVID-19 evolution is investigated, simulating different scenarios defined based on governmental and individual reactions.

## 2. MATHEMATICAL MODEL

A frame-by-frame description of the reality can be represented by a set of differential equations. By assuming only time evolution of state variables, $x \in \Re^n$, spatial aspects are not of concern, allowing to establish a governing equation of the form: $\dot{x} = f(x), x \in \Re^n$. The description of coronavirus disease 2019 dynamics defines its propagation considering animals and humans transmission. Different kinds of populations need to be defined in order to have a proper scenario of the disease propagation.

Lin *et al.* (2020) proposes a susceptible-exposed-infectious-removed (SEIR) framework model to describe the coronavirus disease 2019 (COVID-19). This model was inspired on the original model of He *et al.* (2013) for influenza. Essentially, the description considers a total population of size *N* that contains two classes: *D* is a public perception of risk regarding severe cases and deaths; and *C* represents the cumulative infected cases. In addition, *S* is the susceptible population, *E* is the exposed population, *I* is the infectious population and *R* is the removed population that includes both recovered and deaths. A simplified version of the model considers only person-to-person transmission, and therefore, zoonotic effect is neglected. This scenario assumes the second stage of the Wuhan – China case, after the close of the Huanan Seafood Wholesale Market. Emigration effect is also neglected in order to simplify the original model. Therefore, the simplified version of the governing equations considers the interaction among all these populations, being expressed by the following set of differential equations

$$\dot{S} = -\beta \frac{SI}{N} \tag{1}$$

$$\dot{E} = \beta \frac{SI}{N} - \sigma E \tag{2}$$

$$\dot{I} = \sigma E - \gamma I \tag{3}$$

$$\dot{R} = \gamma_R I \tag{4}$$

$$\dot{D} = d\gamma I - \lambda D \tag{5}$$

$$\dot{C} = \sigma E \tag{6}$$

where the following parameters are defined: $\gamma$ is the mean infectious period; $\gamma_R$ is the adjusted removed period, defining the relation between removed population and the infected one; $\sigma$ is the mean latent period; $d$ is the proportion of severe cases; $\lambda$ is the mean duration of public reaction. It should be pointed out that the parameter $\gamma - \gamma_R$ defines the evolution of the nonreported removed population, which means that, if $\gamma = \gamma_R$, populations are restricted to the classical SEIR case.

The function $\beta = \beta(t)$ represents the transmission rate that considers governmental action, represented by $(1 - \alpha)$; and the individual action, represented by the function $\left(1 - \frac{D}{N}\right)^\kappa$. Therefore, the transmission rate is modeled as follows,

$$\beta = \beta(t) = \hat{\beta}_0 \ (1 - \hat{\alpha})\delta \tag{7}$$

where $\hat{\beta}_0 = \beta_0^{(i)} H\left(t - T_{\beta_0}^{(i)}\right)$ represents the nominal transmission rate and $H\left(t - T_{\beta_0}^{(i)}\right)$ is a step function with the form illustrated in Figure 1. Its use is convenient in order to contemplate variations of the transmission rate through time, being defined as follows.

$$H\left(t - T_{\beta_0}^{(i)}\right) = \begin{cases} \beta_0^{(1)}, & \text{if } t \leq T_{\beta_0}^{(1)} \\ \beta_0^{(2)}, & \text{if } t \leq T_{\beta_0}^{(2)} \\ \beta_0^{(3)}, & \text{if } t \leq T_{\beta_0}^{(3)} \\ \vdots \end{cases} \tag{8}$$

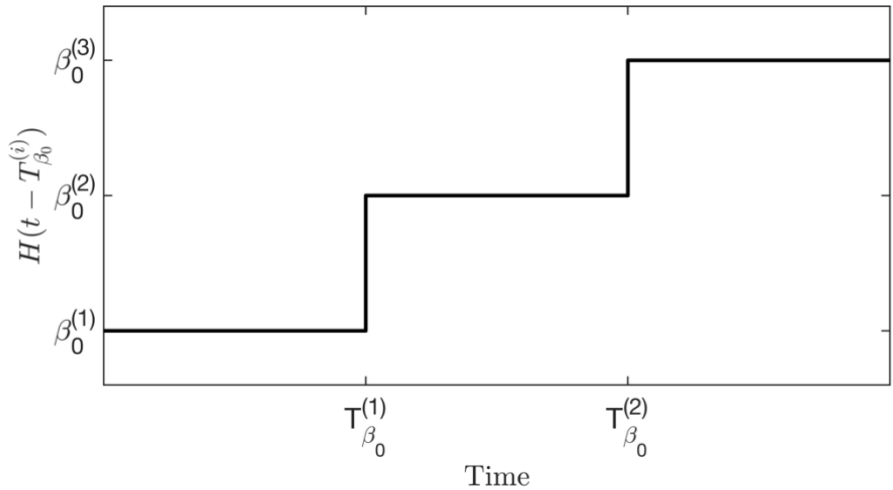

Figure 1: Step function employed to consider parameter variations through time.

Note that, it is assumed that if $T_{\beta_0}^{(m)}$ does not exist, the term $\beta_0^{(m+1)}$ does not exist as well, considering that index $m$ is bigger than 1. Based on that, this general function can represent constant values, or different step functions.

Using the same strategy, it is defined the governmental action as follows:

$$\hat{\alpha} = \alpha_i \, H(t - T_{Gov}^{(i)}) \tag{9}$$

where different steps are considered defined by time instants $T_{Gov}^{(i)}$.

In addition, individual action is represented by

$$\delta = \left(1 - \frac{D}{N}\right)^{\kappa} \tag{10}$$

which the intensity of responses is defined by parameter $\kappa$. These parameters need to be adjusted for each place, being essential for the COVID-19 description.

In general, the parameter definitions depend on several issues, being a difficult task. In this regard, it should be pointed out that real data has spatial aspects that are not treated by this set of governing equations. Hence, this analysis is a kind of average behavior that needs a proper adjustment to match real data. Besides, R Li *et al.* (2020) evaluated Wuhan situation concluding that undocumented novel coronavirus infections are critical for understanding the overall prevalence and pandemic potential of this disease. The authors estimated that 86% of all infections were undocumented and that the transmission rate per person of undocumented infections was 55% of documented infections. This aspect makes the description even more complex.

The use of step functions to define some parameters allows a proper representation of different scenarios, especially the transmission rate. It is also important to observe that either governmental or individual actions have a delayed effect on system dynamics. Virus mutations are another relevant aspect related to the description of coronavirus dynamics that can dramatically alter the system response, but are not treated here.

Numerical simulations are performed considering the fourth-order Runge-Kutta method. The next sections treat the COVID-19 dynamics considering two different objectives. Initially, the next section performed a model verification using information from China, Italy, Iran and Brazil. Afterward, the subsequent section evaluates different scenarios for the Brazilian case, using the parameters adjusted for the verification cases.

## 3. MODEL VERIFICATION

As an initial step of the developed analysis, a model verification is carried out using information available on Worldometer (https://www.worldometers.info/coronavirus/),

considering different countries (Last updates: China – March 26, Italy – Mar 21; Iran – Mar 26; Brazil – Mar 24). The fundamental hypothesis of the analysis is that average populations of the country is of concern. Therefore, it is assumed that each country has a homogeneous distribution, without spatial patterns.

Basically, information from China, Italy, Iran and Brazil are employed. This information is useful to calibrate the model parameters, evaluating its correspondence with real data. Table 1 presents parameters employed for all simulations. They are based on the information of the Lin *et al.* (2020) that, in turn, is based on other references as He *et al.* (2010) and Breto *et al.* (2009). For more details, see other citations referenced therein.

Table 1: Model parameters.

| Parameter | Description | Value |
|---|---|---|
| $\sigma^{-1}$ | Mean latent period | 3 days |
| $\gamma^{-1}$ | Mean infectious period | 5 days |
| $\gamma_R^{-1}$ | Adjusted removed period | 22 days |
| $d$ | Proportion of severe cases | 0.2 |
| $\lambda^{-1}$ | Mean duration of public reaction | 11.2 days |

In addition, susceptible population initial condition is assumed to be $S_0 = 0.9N - E_0 - I_0 - R_0$. In addition, it is assumed that there is no recovered population initially, *i.e.*, $R_0 = 0$. Another information needed for the model is the number exposed persons for each infected person. It is adopted that each infected person has the potential to expose 20 persons, $E_0 = 20I_0$.

Transmission rate considers specific parameters for each case. Nevertheless, the reference values are presented in Table 2.

Table 2: Reference parameters for transmission rate.

| Parameter | Value |
|---|---|
| $\alpha_i$ | [0, 0.4239, 0.8478] |
| $\kappa$ | 1117.3 |

Other parameters are adjusted depending on the case. In the sequence, the dynamics of four different countries is analyzed in order to promote a model verification.

## 3.1 Verification Simulations

The first scenario for the model verification is based on China results. It should be pointed out that this analysis considers all cases in China, not restricted to Wuhan. Parameters presented in Table 3 are employed for simulations with a population of $N = 1.43 \times 10^9$ and an initial state with 554 infected persons ($I_0 = 554$). It should be highlighted again that these parameters are average ones since they are valid for the whole country. Of course, reaction time is different from the distinct parts of the country, which makes necessary to estimate this parameter based on the real data in an average way. Figure 2 presents infected population evolution showing a good agreement between simulation and real data.

Table 3: Model parameters for the transmission rate of China.

| Parameter | Value |
|---|---|
| $\beta_0$ | 0.514 |
| $T_{Gov}^{(i)}$ | [13, 29] days |

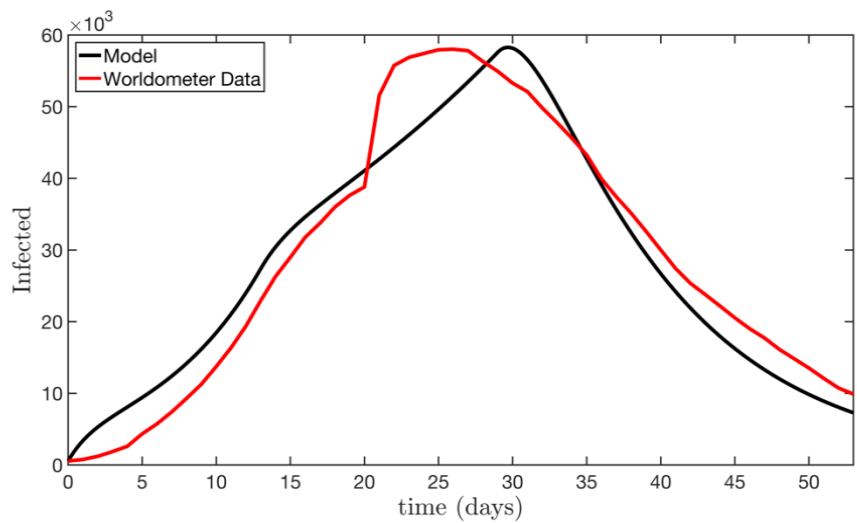

Figure 2: China - infected population through time.

Due to chronological issues, Chinese case is the one with a large number of real data, which makes it useful to establish a comparison of the model prediction error. Figure 3 presents daily errors from china, highlighting the average and maximum errors. Note that the maximum error is less than 28%, with an average error of 13.58%.

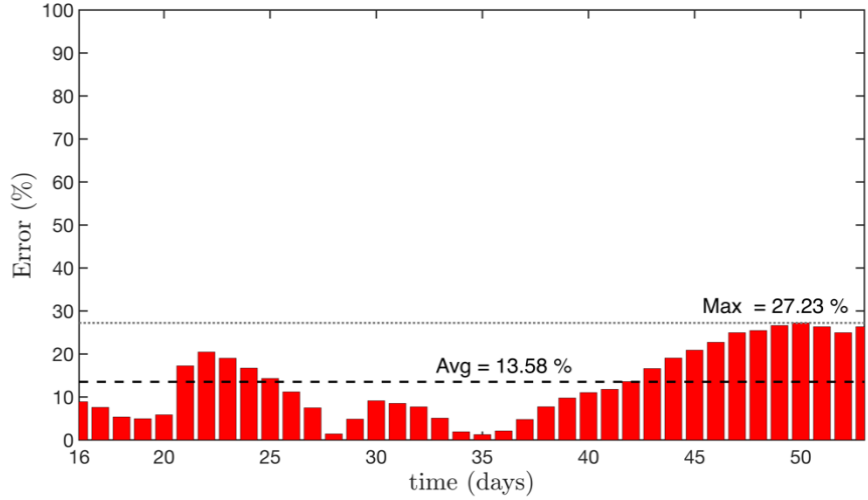

Figure 3: China – prediction errors between the simulated and real data of the infected population.

For the following three cases, Italy, Iran and Brazil, it is assumed that the second stage of governmental action has not been reached yet. Therefore, it is represented by a step function $\alpha_i = [0, 0.4239]$, which means that $T_{Gov}^{(2)}$ is neglected and $\alpha_3$ does not exist.

Italian case is now in focus considering parameters presented in Table 4 with a population of $N = 60.48 \times 10^6$ and an initial state with 20 infected persons ($I_0 = 20$). A step function is considered to define the nominal transmission rate, $\beta_0$, due to extreme governmental actions that have not been effective until present days. Figure 4 presents the infected population simulation compared with real data, showing a good agreement. Figure 5 presents daily errors from Italy, highlighting the average and maximum errors. For this case, the maximum error is less than 19%, with an average error of 10.60%.

Table 4: Model parameters for the transmission rate of Italy.

| Parameter | Value |
| --- | --- |
| $\beta_0^{(i)}$ | [0.594  1.1] |
| $T_{\beta_0}^{(1)}$ | 22 days |
| $T_{Gov}^{(1)}$ | 22 days |

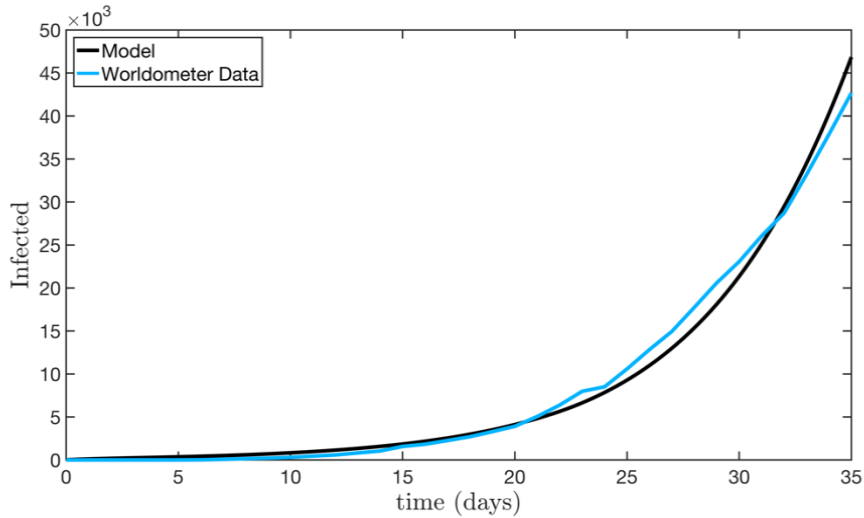

Figure 4: Italy - infected population through time.

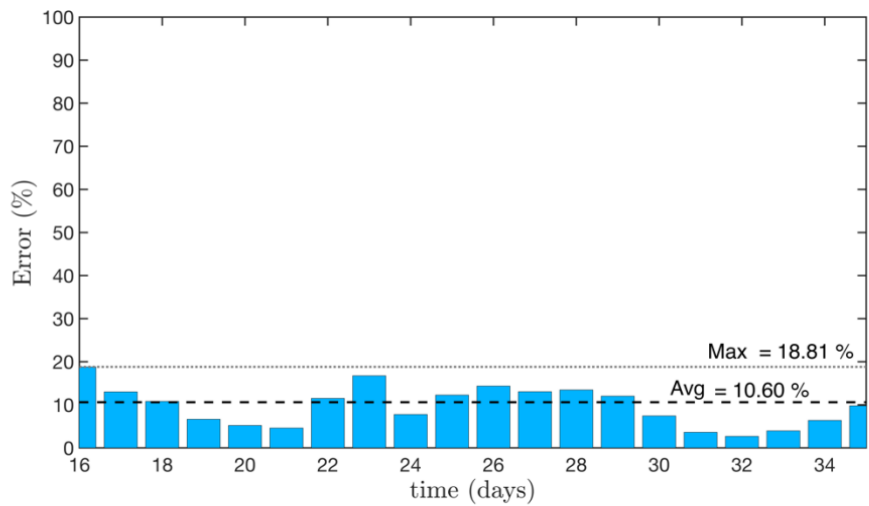

Figure 5: Italy – prediction errors between the simulated and real data of the infected population.

Iran case is now treated considering the parameters presented in Table 5 with a population of $N = 81.16 \times 10^6$ and an initial state with 20 infected persons ($I_0 = 20$). Results are presented in Figure 6 showing a good agreement with real data. Figure 7 presents daily errors, highlighting the average and maximum errors. Although the average error is 15.46%, the maximum error is around 42%, which is a large value. Nevertheless, it should be observed that the big values are related to the beginning of the predictions, probably due to problems with the original data.

Table 5: Model parameters for the transmission rate of Iran.

| Parameter | Value |
|---|---|
| $\beta_0^{(1)}$ | 0.594 |
| $T_{Gov}^{(1)}$ | 24 days |

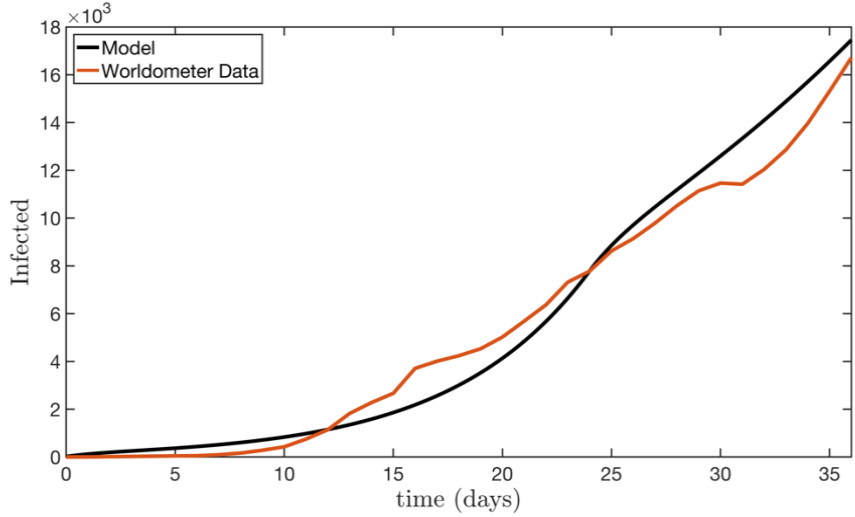

Figure 6: Iran - infected population through time.

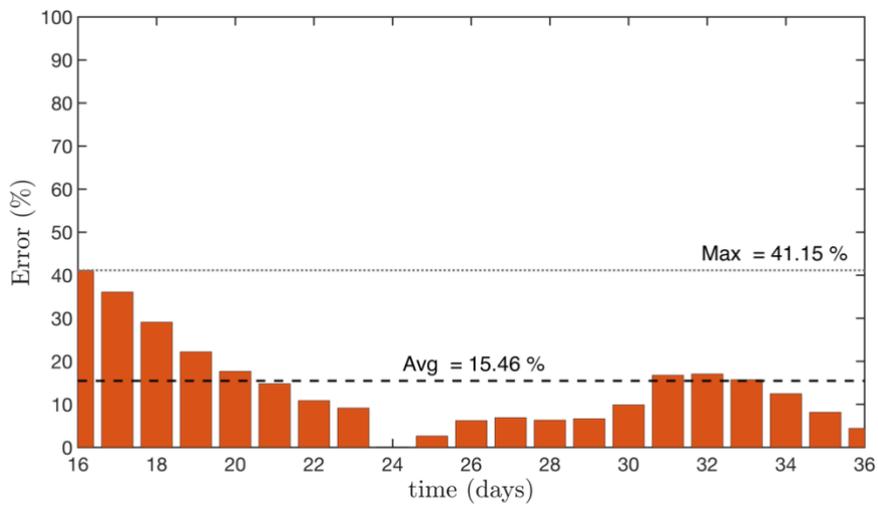

Figure 7: Iran – prediction errors between the simulated and real data of the infected population.

Brazilian case is now of concern considering parameters presented in Table 6 with a population of $N = 209.3 \times 10^6$ and an initial state with 10 infected person ($I_0 = 10$). Figure 8 presents the infected population evolution showing that the same trend of the other cases is followed, being enough to have a general scenario. It should be highlighted that Brazilian outbreak is in the beginning, with information that is not enough for a better calibration.

Table 6: Model parameters for the transmission rate of Brazil.

| Parameter | Value |
|---|---|
| $\beta_0^{(1)}$ | 0.675 |
| $T_{Gov}^{(1)}$ | 17 days |

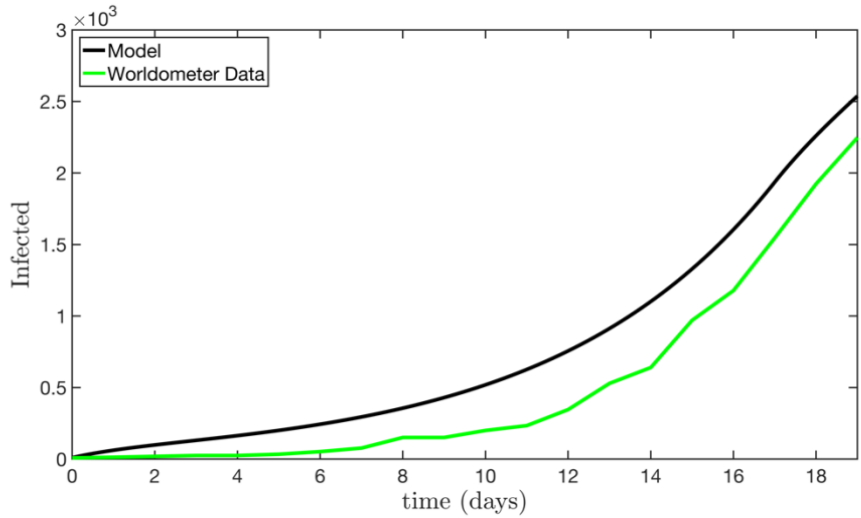

Figure 8: Brazil - infected population through time.

## 4. BRAZILIAN SCENARIOS

This section has the objective to investigate different scenarios related to COVID-19 dynamics in Brazil. Parameters adjusted on the previous section are employed to evaluate different scenarios varying governmental and individual reactions. It should be pointed out that this adjustment does not have enough information, but it is possible to perform, at least a qualitative analysis of the COVID-19 dynamics in Brazil.

Initially, two different transmission rates are defined: naive scenario, without intervention ($\alpha = \kappa = 0$); and with governmental and individual actions ($\alpha \neq 0; \kappa \neq 0$). Figure 9 presents numerical simulations together with the real data that is presented just for the first days. The same parameters presented in Table 6 are employed assuming $T_{Gov}^{(2)} = 37$ days. A logarithm scale is adopted since the naive scenario has a dramatic increase of the infected cases. Besides the big difference between both cases, it is clear the huge impact of variations on the transmission rate function that represents governmental and individual actions. It is noticeable that the effective actions tend to reduce the infected population, reducing the final crisis period as well.

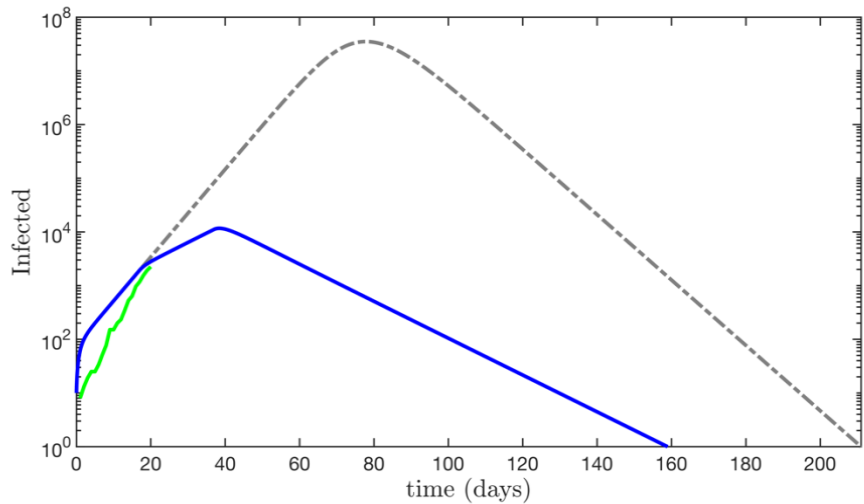

Figure 9: Transmission evolution considering two different scenarios for different transmission rates: naive scenario, without intervention ($\alpha = \kappa = 0$); and with governmental intervention and individual action with the values adjusted in the previous section.

A more detailed analysis of the COVID-19 dynamics is treated considering the other populations for the case with intervention treated in Figure 9 (parameters of Table 6 with $T_{Gov}^{(2)} = 37$ days). Figure 10 presents all system state variables, showing the susceptible, *S*, exposed, *E*, infected, *I*, removed, *R*, public perception, *D*, and cumulative cases, *C*. The interaction among all the populations defines a kind of equilibrium established by the governing equations.

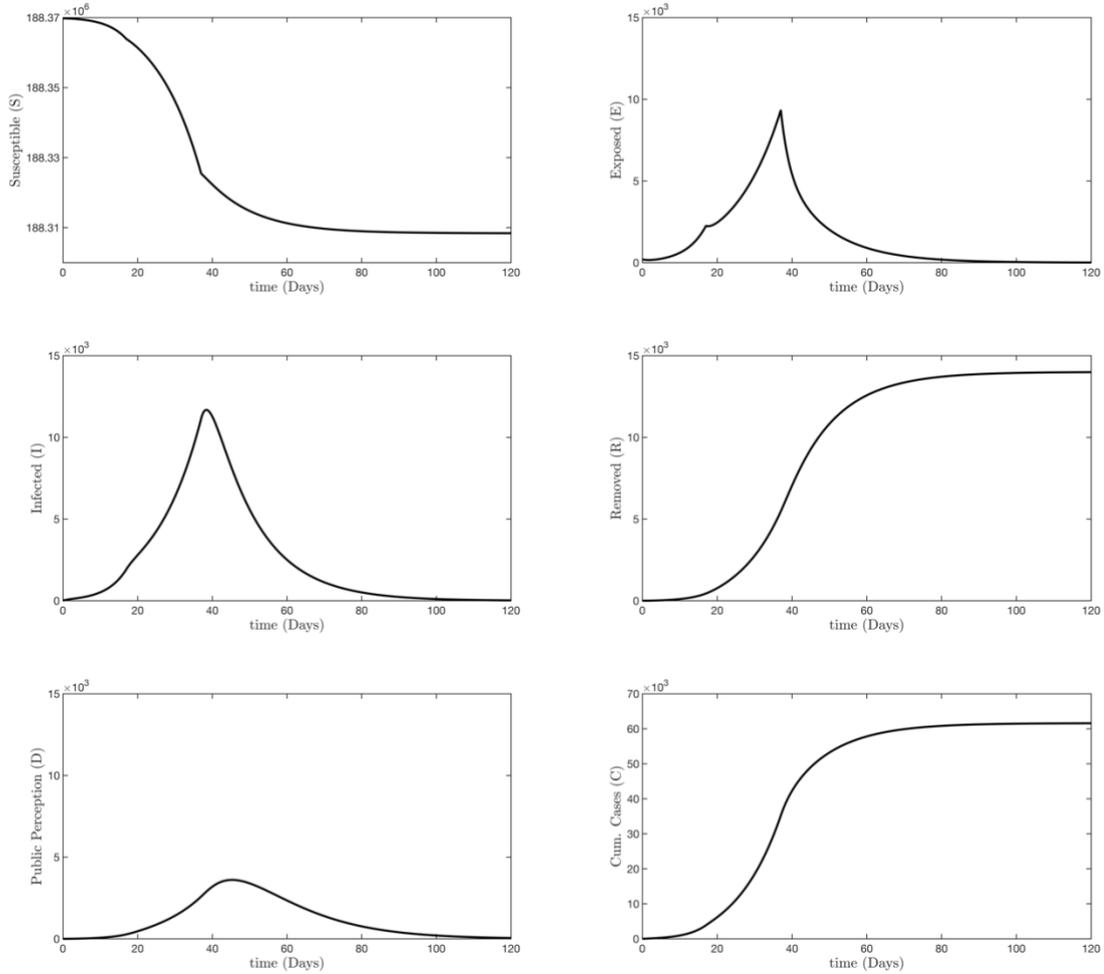

Figure 10: Population interactions considering a scenario with intervention: susceptible, *S*, exposed, *E*, infected, *I*, removed, *R*, public perception, *D*, and cumulative cases, *C*.

Nowadays, one of the most relevant issue to be discussed in terms of propagation is the governmental and individual actions. A parametric analysis is of concern considering distinct scenarios related to intervention. Scenarios defined by the variation of the intervention moments is initially treated. The moment of the governmental action start, represented by parameter $T_{Gov}^{(1)}$ (day), is analyzed in Figure 11, considering the following values: 17, 22, 27 and 32 and $T_{Gov}^{(2)}$ is assumed to be 20 days after $T_{Gov}^{(1)}$. Note that the delay to the start of the governmental action dramatically alters the response, increasing the number infected population and its duration. The same conclusion can be established considering the second governmental action, represented by $T_{Gov}^{(2)}$ (day), presented in Figure 12 that shows the same trend considering a different set of start instants: 37, 42, 47 and 52.

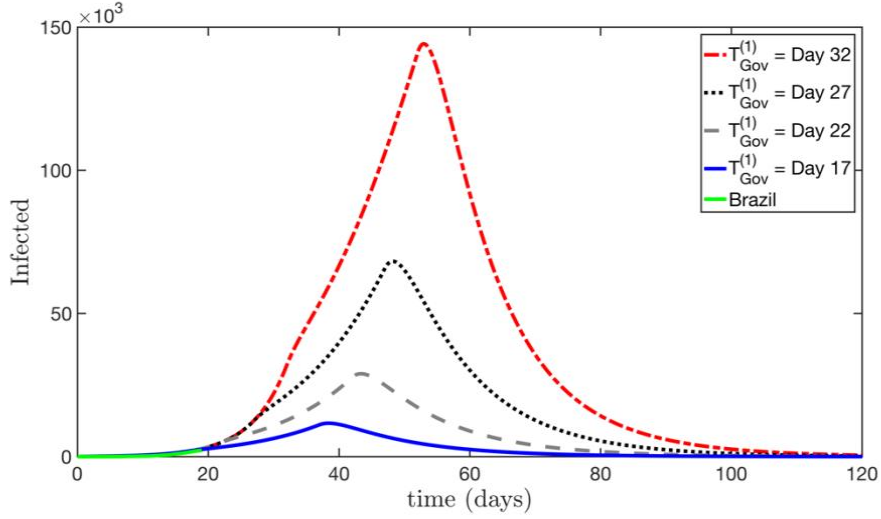

Figure 11: Infected population evolution considering different first governmental action start, represented by parameter $T_{Gov}^{(1)}$ (day): 17, 22, 27 and 32.

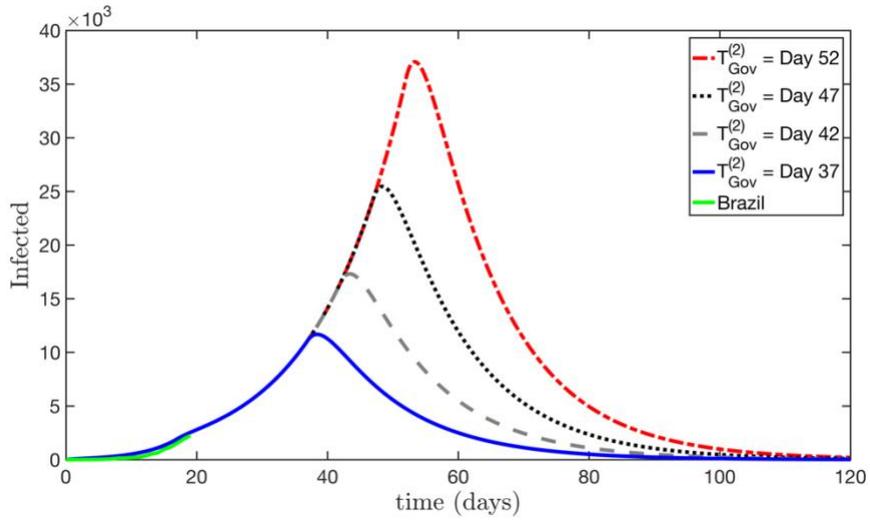

Figure 12: Infected population evolution considering different second governmental action start, represented by parameter $T_{Gov}^{(2)}$ (day): 37, 42, 47 and 52.

A scenario with a governmental action that starts, finishes and then restarts again is now evaluated, considering the following parameters: $T_{Gov}^{(i)} = [17, 37, 52]$ and $\alpha_i = [0, 0.4239, 0, 0.8478]$. This scenario is compared with the usual one where the intervention starts in a level and then evolve to a more severe situation, considering: $T_{Gov}^{(i)} = [17, 52]$ and $\alpha_i = [0, 0.4239, 0.8478]$. Figure 13 shows both situations represented by the transmission function and the evolution of infected populations. It is clear that the interruption of the governmental action causes a dramatic worst scenario.

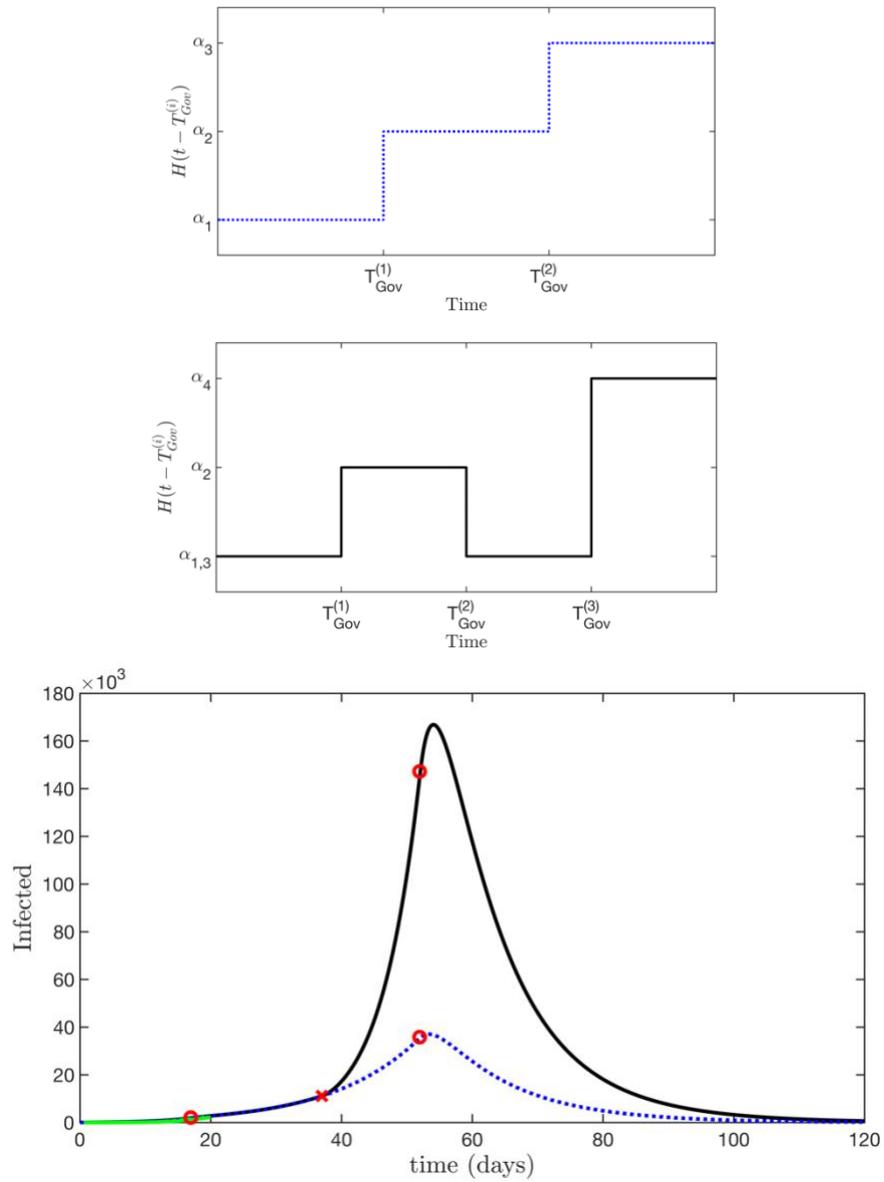

Figure 13: Infected population evolution considering different scenarios related to the governmental action, $\hat{\alpha}$.

## 5. CONCLUSIONS

A mathematical model based on the susceptible-exposed-infectious-recovered framework is employed to describe the COVID-19 evolution. A verification procedure is performed based on the available data from China, Italy, Iran and Brazil. Afterward, different scenarios from Brazil is analyzed. Results clearly show that governmental and individual actions are essential to reduce the infected populations and also the total period of the crisis. The mathematical model can be improved in order to include more phenomenological information that can increase its capability to describe different scenarios. Nevertheless, it should be pointed out that the mathematical model and its numerical simulations are important tools that can be useful for public health planning.

## 6. ACKNOWLEDGEMENTS

The authors would like to acknowledge the support of the Brazilian Research Agencies CNPq, CAPES and FAPERJ. Since this work was developed during a quarantine period, the authors would like to express their gratitude to familiar support that, besides the patience, helps to collect available information. Therefore, it is important to acknowledge: Raquel Savi, Rodrigo Savi, Antonio Savi and Bianca Zattar.

- Li Q, Med M, Guan X, Wu P, Wang X, Zhou L et al. (2020), "Early transmission dynamics in Wuhan, China, of novel coronavirus-infected pneumonia", *The New England Journal of Medicine*, doi:10.1056/NEJMoa2001316.
- Lin Q, Zhao S, Gao D, Lou Y, Yang S, Musa SS, Wang MH, Cai Y, Wang W, Yang L, Hee D (2020), "A conceptual model for the coronavirus disease 2019 (COVID-19) outbreak in Wuhan, China with individual reaction and governmental action", *International Journal of Infectious Diseases*, v.93, pp.211–216.
- Li R, Pei S, Chen B, Song Y, Zhang T, Yang W & Shaman J (2020), "Substantial undocumented infection facilitates the rapid dissemination of novel coronavirus (SARS-CoV2)", *Science*, doi:10.1126/science.abb3221.
- Rihan FA, Al-Salti NS & Anwar M-NY (2018), "Dynamics of coronavirus infection in human", *AIP Conference Proceedings*, v.1982, Article 020009, doi:10.1063/1.5045415.
- Riou J & Althaus CL (2020), "Pattern of early human-to-human transmission of Wuhan 2019 novel coronavirus (2019-nCoV), December 2019 to January 2020", *Euro Surveill*, v.25, n. 4, Article 2000058. doi:10.2807/1560-7917.ES.2020.25.4.2000058.
- Savi MA (2005), "Chaos and order in biomedical rhythms", *Journal of the Brazilian Society of Mechanical Sciences and Engineering*, v.27, n.2, pp.157-169.
- Wu JT, Leung K, Leung GM (2020), "Nowcasting and forecasting the potential domestic and international spread of the 2019-nCoV outbreak originating in Wuhan, China: a modelling study", *Lancet*, v.395, pp.689-697.doi:10.1016/ S0140-6736(20)30260-9.